\documentclass[12pt,preprint]{aastex}
\usepackage{amsmath,amssymb}

\newcommand{\exponint}[2]
{\mathrm{E}_{#1}(#2)}

\begin{document}

\title{Influence of atomic polarization and horizontal illumination on the
Stokes profiles of the He {\sc i} 10830\AA\ multiplet}

\author{Javier Trujillo Bueno\altaffilmark{1,2,3} and Andr\'es Asensio Ramos \altaffilmark{1}}
\altaffiltext{1}{Instituto de Astrof\'{\i}sica
de Canarias, 38205, La Laguna, Tenerife, Spain} 
\altaffiltext{2}{Institut f\"ur Astrophysik, Friedrich-Hund-Platz 1, 37077 G\"ottingen, Germany}
\altaffiltext{3}{Consejo Superior de Investigaciones Cient\'{\i}ficas,
Spain} \email{{\bf (The Astrophysical Journal; in press)}}

\begin{abstract}

The polarization observed in the spectral lines of the He {\sc i} 10830 \AA\ multiplet carries valuable information on the dynamical and magnetic properties of plasma structures in the solar chromosphere and corona, such as spicules, prominences, filaments, emerging magnetic flux regions, etc. Therefore, it is crucial to have a good physical understanding of the sensitivity of the observed spectral line polarization to the various competing physical mechanisms. Here we focus on investigating the influence of atomic level polarization on the emergent Stokes profiles for a broad range of magnetic field strengths, in both $90^{\circ}$ and forward scattering geometry. We show that, contrary to a widespread belief, the selective emission and absorption processes caused by the presence of atomic level polarization may have an important influence on the emergent linear polarization, even for magnetic field strengths as large as 1000 G. Consequently, the modeling of the Stokes $Q$ and $U$ profiles should not be done by taking only into account the contribution of the transverse Zeeman effect within the framework of the Paschen-Back effect theory, unless the magnetic field intensity of the observed plasma structure is sensibly larger than 1000 G. We point out also that in low-lying optically thick plasma structures, such as those of active region filaments, the (horizontal) radiation field generated by the structure itself may substantially reduce the positive contribution to the anisotropy factor caused by the (vertical) radiation field coming from the underlying solar photosphere, so that the amount of atomic level polarization may turn out to be negligible. Only under such circumstances may the emergent linear polarization of the He {\sc i} 10830 \AA\ multiplet in such regions of the solar atmosphere be dominated by the contribution caused by the transverse Zeeman effect. 

\end{abstract}


\section{Introduction}

In order to obtain reliable empirical information on the strength and geometry
of astrophysical magnetic fields we need  to develop and apply suitable
diagnostic tools within the framework of the quantum theory of spectral line
polarization \citep[e.g., the recent monograph by][]{landi_landolfi04}. 
A na\"ive knowledge and/or an unjustified neglect of some of the physical
mechanisms that produce polarization in spectral lines may lead to significant
errors in our inferences and/or to missing the exciting opportunity of proposing
new practical ideas for magnetic field diagnostics. In this respect, the main
aim of this paper is to help clarify further the physics of the observed
polarization in the spectral lines of the He {\sc i} 10830 \AA\ multiplet,
because they contain precious information on the dynamical and magnetic
properties of a variety of solar chromospheric and coronal structures.

The most important mechanisms that induce and modify polarization signatures in
the spectral lines that originate in the atmospheres of the
Sun and of other stars are the Zeeman effect, anisotropic radiation pumping and the Hanle effect \citep[e.g., the recent review by][]{trujillo_05}. 

The Zeeman effect requires the presence of a magnetic field, which 
causes the atomic and molecular energy levels to split
into different magnetic sublevels characterized by their magnetic quantum number
$M$. As a result, the wavelength positions of the 
$\pi$ ($\Delta{M}=M_u-M_l=0$), $\sigma_{\rm blue}$ ($\Delta{M}=+1$) and $\sigma_{\rm
red}$ ($\Delta{M}=-1$) transitions do not coincide and, therefore, their
respective polarization signals do not cancel out. The Zeeman effect is most
sensitive in {\em circular} polarization (quantified by the Stokes $V$
parameter), 
with a magnitude that for not too strong fields scales with the ratio 
between the Zeeman splitting
and the width of the spectral line (which is very much larger than
the natural width of the atomic levels!). 
This so-called {\em longitudinal} Zeeman effect 
responds to the line-of-sight component of the magnetic field.
In contrast, the {\em transverse} Zeeman effect
responds to the component of the magnetic field perpendicular
to the line of sight, producing instead {\em linear} polarization 
signals (quantified by the Stokes $Q$ and $U$ parameters)
that are normally below the noise level of present observational
possibilities for intrinsically weak fields (typically B${\lesssim}100$ gauss for solar spectropolarimetry).

The anisotropic illumination of the atoms 
in the outer regions of a stellar atmosphere may produce
atomic level polarization (that is, population imbalances and/or quantum
coherences between the magnetic sublevels pertaining to the upper and/or lower
level of the line transition under consideration), in 
such a way that the populations of substates with different values of $|M|$
are different. This is termed {\em atomic level alignment}.
Under such circumstances, the polarization signals of the $\pi$ and $\sigma$ transitions do
not cancel out, even in the absence of a magnetic field, simply because the
population imbalances among the magnetic sublevels imply more or less $\pi$ 
transitions, per unit volume and time, than $\sigma$ transitions. Interestingly,
this atomic level polarization is modified by the presence of a magnetic field
{\em inclined} with respect to the symmetry axis of the pumping radiation field
\citep[i.e., the Hanle effect; e.g., the review by][]{trujillo01}. The magnetic field
intensity (measured in gauss) that is sufficient to produce a sizable change in the atomic polarization of a given level is

\begin{equation}
B_H\,=\, \frac{1.137 \times 10^{-7}}{t_\mathrm{life} \, g_{J}}\, ,
\end{equation}
where $t_{\rm life}$ and $g_{J}$ are, respectively, the level's lifetime (in
seconds) and its Land\'e factor\footnote{This basic formula of the Hanle effect
results from equating the Zeeman splitting with the natural width (or inverse
lifetime) of the energy level under consideration (which can be either the upper
or the lower level of the chosen spectral line).}.

As discussed by \cite{trujillo_nature02} a 
notable example of a multiplet whose spectral lines are sensitive to both
effects (that is, to the Zeeman splitting and to the atomic level polarization)
is the He {\sc i} 10830 \AA\ multiplet, whose Stokes profiles can be observed
in a variety of plasma structures of the solar chromosphere and corona, such as
sunspots \citep{harvey_hall71,ruedi95,centeno06},
coronal filaments \citep{lin98,trujillo_nature02}, prominences
\citep{trujillo_nature02,merenda06}, emerging flux regions
\citep{solanki_nature03,Lagg04}, chromospheric spicules 
\citep{trujillo_merenda05,socas_elmore05}, active region filaments (Mart\'\i
nez Pillet et al. 2006; in preparation) and flaring regions (Sasso et al. 2006).
 
The He~{\sc i} 10830 \AA\ multiplet originates between a lower term ($2^3{\rm
S}_1$) and an upper term ($2^3{\rm P}_{2,1,0}$). Therefore, it comprises three
spectral lines: a `blue' component at 10829.09~\AA\ (with $J_l=1$ and $J_u=0$),
and two `red' components at 10830.25~\AA\ (with $J_u=1$)
and at 10830.34~\AA\ with ($J_u=2$) which appear 
blended at solar atmospheric temperatures. 
Although the circular polarization of the He {\sc i} 10830 \AA\ lines is
dominated by the longitudinal Zeeman effect, we expect their linear
polarization to be the result of the joint action of the transverse Zeeman effect
and of the atomic polarization that anisotropic radiation pumping processes
induce in the helium levels.
Actually, as we shall show in this paper, the degree of anisotropy of the solar
continuum radiation at the wavelengths of the relevant helium transitions is
sufficiently important so as to produce a significant amount of atomic
polarization in the lower and upper levels of the He {\sc i} 10830 multiplet,
even at relatively low atmospheric heights  (e.g., ${\sim}$1000 Km). Elastic
collisions with neutral hydrogen atoms are unable to destroy this atomic
polarization that the anisotropy of the solar radiation field can (in principle)
induce in the helium levels. Obviously, for sufficiently weak magnetic fields
(e.g., for $B{\lesssim}100$ G) the linear polarization is dominated by the {\em
selective emission} and {\em selective absorption} of polarization components
that result from the atomic level polarization \citep{trujillo_nature02}.
However, it would not be correct to give for granted that for relatively strong
fields (say, for $B{\approx}1000$ G) the linear polarization of the He {\sc i}
10830 \AA\ multiplet is going to be necessarily dominated by the transverse
Zeeman effect. 

Unfortunately, with very few exceptions \citep{trujillo_nature02,trujillo_merenda05,merenda06}, 
the modeling of spectropolarimetric observations of the He
{\sc i} 10830 \AA\ multiplet has been carried out without taking into account
that, in principle, the observed linear polarization may be actually the result
of the joined action of the transverse Zeeman effect and of the atomic level
polarization. To neglect {\em a priori} the influence of atomic level
polarization simply because the considered point of the observed field of view
is significantly magnetized is not justified. This is
a reasonable approximation when the plasma diagnostics is being done via the
analysis of only the Stokes $I$ and $V$ profiles, such as those measured in
sunspot umbrae \citep[e.g.,][]{centeno06}, but 
it might not be so adequate in the case that Stokes $Q$ and $U$ are also considered, such as those observed by \cite{solanki_nature03}
and \cite{Lagg04} in emerging flux regions. 

A recent investigation 
has already pointed out that, even when only the contribution of the Zeeman
effect is accounted for, the wavelength positions and the strengths of the $\pi$
and $\sigma$ components should be calculated in the incomplete Paschen-Back
effect regime, given that the linear Zeeman effect theory {\em overestimates}
the amplitudes of the emergent linear and circular polarization --that is, it
{\em underestimates} the inferred magnetic field strength \citep{socas_navarro04}. 
In this paper we focus instead on clarifying the
role played by the presence of atomic level polarization on the emergent Stokes
profiles of the He {\sc i} 10830 \AA\ multiplet for increasing values of the
strength of the assumed magnetic field. As we shall see below, the influence of
atomic level polarization on the Stokes $Q$ and $U$ profiles of the He {\sc i}
10830 \AA\ multiplet turns out to be significant, even for 
magnetic field strengths as large as 1000 G. In fact, as pointed out below, some of the 
Stokes profiles of emerging flux regions that \cite{solanki_nature03}
and \cite{Lagg04} have interpreted in terms of the linear Zeeman effect theory
neglecting atomic level polarization
show, however, clear observational signatures of the presence of this physical
ingredient. On the other hand, if any Stokes profiles observation of a
moderately magnetized solar atmospheric region turns out to show no hint at all
of the presence of atomic level polarization (such as it seems to be the case
with the active region filament observations reported by V. Mart\'\i nez Pillet
and collaborators during the Fourth International Workshop on Solar
Polarization) the reason, in our opinion, is to be found in the lower degree of anisotropy of the radiation field that pumps the helium atoms inside such elongated, optically-thick plasma structures. 

The outline of this paper is as follows. The formulation of the problem is
presented in Section 2, where we inform about the equations we have solved for
calculating the emergent Stokes profiles from an optically thick slab located at
a height of about 2000 Km above the solar visible ``surface". Section 3 shows several
illustrative examples of the generated spectral line polarization when
neglecting or taking into account the atomic level polarization induced by the photospheric continuum radiation. Section 4 investigates the extent to which the anisotropy of the photospheric radiation field may be modified by the radiation generated by the assumed plasma structure itself. Finally, in Section 5 we summarize the main conclusions of our work and insist on the scientific interest for spectropolarimetry from space.

\section{Formulation of the problem}

As illustrated in Fig. 1, we consider a constant-property slab of magnetized
chromospheric material located at a height of only 3 arc seconds (${\sim}2000$) km above the solar
``visible" surface. 
The magnetic field, whose strength we will vary at will, is
assumed to be horizontal (i.e., parallel to the solar ``surface"). The slab's
optical thickness at the wavelength and line of sight under consideration is
$\tau$, while 
the symbol $\Delta{\tau}_{\rm red}$ denotes the optical thickness of the slab
along its normal direction at the line center of the red blended component. We
have chosen $\Delta{\tau}_{\rm red}=1$, which is just at the transition limit
between an optically thin and an optically thick medium. All the
atoms inside this slab are assumed to be illuminated by the unpolarized and
limb-darkened photospheric radiation field whose center-to-limb variation has
been tabulated by \cite{cox00}. This implies that, for the moment, we are
neglecting the influence of possible radiative transfer effects inside this
$\Delta{\tau}_{\rm red}=1$ slab on the anisotropy factor, whose definition is
\citep{landi_landolfi04}

\begin{equation}
w=\sqrt{2}\frac{J^2_0}{J^0_0}\,,
\end{equation}
where 
\begin{equation}
J^0_0=\frac{1}{4\pi}\int{I_{\nu,{\vec\Omega}}d{{\vec\Omega}}}
\end{equation}
and
\begin{equation}
J^2_0=\frac{1}{2\sqrt{2}}
\frac{1}{4\pi}\int{(3{\mu}^2\,-1\,)I_{\nu,{\vec\Omega}}d{{\vec\Omega}}},
\end{equation}
with $\mu={\rm cos}\,{\theta}$ ($\theta$ being the angle between the ray under consideration and the solar local vertical). Therefore, in a stellar atmosphere the possible values of the anisotropy factor vary between $w=-1/2$ for the limiting case of illumination by a purely horizontal radiation field without any azimuthal dependence, and $w=1$ for purely vertical illumination.
Two main factors contribute to the value of $w$ at a given height in the {\em
quiet} solar atmosphere: (1) the center to limb variation of the solar radiation
field at the wavelength under consideration and (2) the geometrical effect due
to the fact that, the larger the atmospheric height, the smaller the solid angle
subtended by the solar visible sphere. We point out that, at a height of
3 arc seconds in the quiet solar atmosphere, $w=0.097$ at 10830 \AA, while $w=0.040$ if only the
above-mentioned geometrical effect is taken into account (that is, when the contribution due to the center-to-limb variation is disregarded).

The radiative transitions caused by the
anisotropic illumination of the slab's helium atoms 
induce population imbalances and quantum coherences among the magnetic
substates of the energy levels (that is, atomic level polarization), which we quantify
by solving the statistical equilibrium equations for the multipole components,
$\rho^K_Q(J,J^{'})$, of the atomic density matrix \citep[see Section 7.6a in][]{landi_landolfi04}. We do this using a realistic model atom that includes the 5 lowest terms of the triplet system of neutral helium, which implies 11 $J$-levels and 6 transitions between the terms (see Fig. 13.9 in Landi Degl'Innocenti \& Landolfi 2004).  
Such equations take fully into account the Hanle and Zeeman effects produced by the
assumed horizontal magnetic field. We point out that we calculate the wavelength
positions and the strengths of the Zeeman components in the incomplete
Paschen-Back effect regime. From the calculated density-matrix elements it is
then possible to compute the coefficients $\epsilon_I$ and $\epsilon_X$ (with
$X=Q,U,V$) of the emission vector and the coefficients $\eta_I$, $\eta_X$, and
$\rho_X$ of the $4\times4$ propagation matrix of the Stokes vector transfer
equation for a wavelength interval covering the 10830 \AA\ multiplet
\citep[see Sections 7.6b in][]{landi_landolfi04}.

The emergent Stokes vector $\mathbf{I}(\nu,\mathbf{\Omega})=(I,Q,U,V)^{\dag}$
(with $\dag$=transpose, $\nu$ the frequency and $\mathbf{\Omega}$ the line of
sight) is given by the following expression, which can be easily obtained as a
particular case of Eq. (27) of \cite{trujillo03}:

\begin{equation}
\mathbf{I} = \left[ \mathbf{1}+\Psi_0 \mathbf{K}' \right]^{-1} \left[ \left(
e^{-\tau} \mathbf{1} - \Psi_M \mathbf{K}' \right) \mathbf{I}_{\rm sun} +
(\Psi_M+\Psi_0) \mathbf{S} \right],
\label{eq:shortcar}
\end{equation}
where $\mathbf{1}$ is the identity matrix and $\mathbf{I}_{\rm sun}$  the Stokes
vector that illuminates the slab's boundary that is most distant from the
observer, while $\mathbf{K}'$ and $\mathbf{S}$ are given by
\begin{eqnarray}
\mathbf{K}' &=& \frac{\mathbf{K}}{\eta_I}-\mathbf{1}, \\
\mathbf{S} &=& \frac{\boldsymbol{\epsilon}}{\eta_I}.
\end{eqnarray}
In these expressions $\mathbf{K}$ is the propagation matrix of the Stokes vector
transfer equation (whose elements are ${\eta_I}$, ${\eta_X}$ and ${\rho_X}$; $X$
being $Q, U$ or $V$), while
$\boldsymbol{\epsilon}=(\epsilon_I,\epsilon_Q,\epsilon_U,\epsilon_V)^{\dag}$ is
the emission vector.
The coefficients $\Psi_M$ and $\Psi_0$ depend only on the optical depth 
of the slab at the frequency and line-of-sight under consideration
and their expressions are:
\begin{eqnarray}
\Psi_M&=& \frac{1-e^{-\tau}}{\tau} - e^{-\tau},\nonumber \\
\Psi_0 &=&1-\frac{1-e^{-\tau}}{\tau}.
\end{eqnarray}

It is of interest to note that when the anomalous dispersion terms are neglected
in Eq. (\ref{eq:shortcar}) (i.e., the $\rho_X$ terms are taken equal to zero) we obtain

\begin{equation}
I({\tau})\,=\,I_0\,{\rm e}^{-{\tau}}\,+\,{\frac{\epsilon_I}{\eta_I}}(1\,-\,{\rm
e}^{-{\tau}}),
\label{eq:stokesI}
\end{equation}

\begin{equation}
X({\tau})\,=\,X_0\,{\rm e}^{-{\tau}}\,+\,{\frac{\epsilon_X}{\eta_I}}(1\,-\,{\rm
e}^{-{\tau}})\,-\,
{\frac{{\epsilon_I}{\eta_X}}{{\eta_I}^2}}(1\,-\,{\rm e}^{-{\tau}})\,+\,
{\frac{\eta_X}{\eta_I}}{{\tau}}{\rm
e}^{-{\tau}}({\frac{\epsilon_I}{\eta_I}}\,-\,I_0).
\label{eq:stokesX}
\end{equation}
These approximate formulae for the emergent Stokes parameters coincide with
those 
proposed by \cite{trujillo_merenda05} for modeling the Hanle and Zeeman
effects in solar chromospheric spicules and coronal filaments, which provide a
very good approximation whenever $\epsilon_I{\gg}\epsilon_X$ and
$\eta_I{\gg}(\eta_X,\rho_X)$. 
Although such inequalities are often met in solar spectropolarimetry 
\citep[e.g.,][]{sanchez_almeida99},
all the calculations of this paper have however been carried out using the exact
analytical solution given by Eq. (\ref{eq:shortcar}), because they allow us to provide accurate
results also for relatively high field strength values.

It is of interest to point out that Eqs. (\ref{eq:stokesI}) and (\ref{eq:stokesX}) simplify as follows for
the optically thin and optically thick limiting cases:

(1) {\em Optically thin case} ($\tau{\ll}1$)

\begin{equation}
I({\tau})\,{\approx}\,I_0\,+\,{\tau}(S_I\,-\,I_0)
\end{equation}

\begin{equation}
X({\tau})\,{\approx}\,X_0\,(1\,-\,{\tau})\,+\,{\tau}\big{(}S_I\,{\frac{\epsilon_X}{
\epsilon_I}}\,
-\,I_0{\frac{\eta_X}{\eta_I}}\big{)}.
\end{equation}

(2) {\em Optically thick case} ($\tau{\gg}1$)

\begin{equation}
I({\tau})\,{\approx}\,S_I
\end{equation}

\begin{equation}
X({\tau})\,{\approx}\,S_I\,\big{(}{\frac{\epsilon_X}{\epsilon_I}}\,
-\,{\frac{\eta_X}{\eta_I}}\big{)},
\end{equation}
where $S_I={\frac{\epsilon_I}{\eta_I}}$.

\section{The influence of atomic polarization}

This section presents results for the two line of sights illustrated in Fig. 1: $\mu=0$ and
$\mu=1$, where $\mu={\rm cos}\,{\theta}$ ($\theta$ being the angle between the
solar radius vector through the observed point and the line of sight).
Therefore, $\mu=0$ corresponds to the case of an off-limb observation
($90^{\circ}$ scattering geometry), while $\mu=1$ to that of a solar disk center
observation (forward scattering geometry). We point out that 
the boundary conditions are $I_0=X_0=0$ for the $90^{\circ}$ scattering case,
but $I_0=I_{\rm sun}(\mu=1)$ and $X_0=0$ for the forward scattering case (with $I_{\rm sun}(\mu=1)$ taken from \cite{cox00}).

Figure \ref{fig:fig2} shows the emergent Stokes profiles corresponding to the two lines of sight illustrated in Fig. 1.
The left panels of Fig. 2 concern the $90^{\circ}$ scattering or $\mu=0$ case,
which is typical of any off-limb observation. The right panels consider the
forward scattering or $\mu=1$ case, which is typical of an on-disk observation at or close to the
center of the solar disk. In both cases we have assumed a constant-property slab
with $\Delta{\tau}_{\rm red}=1$ located at a height of only 3 arc seconds above
the visible solar surface. As mentioned above, the magnetic field is assumed to be horizontal (i.e.,
parallel to the solar ``surface") and oriented as indicated in Fig. 1 -- that
is, in a way such that also for the off-limb case the magnetic field vector is
perpendicular to the line of sight. From top to bottom Fig. \ref{fig:fig2} shows the emergent
Stokes profiles for increasing values of the magnetic strength and for the
following three calculations of increasing realism:

(1) The dotted lines indicate the case without atomic level polarization. Here
the only mechanism responsible for the emergent polarization is the Zeeman
splitting of the upper and/or lower energy levels, which produces wavelength shifts
between the $\pi$ and $\sigma$ components, whose positions and strengths have been
calculated in the incomplete Paschen-Back effect regime. Therefore, zero
polarization is found for $B=0$ G.

(2) The dashed lines correspond to the case in which we have taken into account
the influence of the atomic polarization of the two upper levels of the He {\sc
i} 10830 \AA\ multiplet that can carry atomic level polarization -that is, those
with $J=2$ and $J=1$. Therefore, in addition to the above-mentioned Zeeman
effect contribution, we have here the possibility of a selective emission of
polarization components, even for the zero field case. For example, this is the
reason why the dashed line of the upper left panel of Fig. \ref{fig:fig2} shows a non-zero
linear polarization signal for the off-limb zero field case.

(3) The solid lines correspond to the most general situation in which the
influence of the atomic polarization of the lower level is also taken into
consideration, in addition to that of the upper levels and to the Zeeman effect.
The consideration of lower-level polarization has two consequences. First, the
amount of upper level polarization and the ensuing selective emission of
polarization components is modified. Second, we can also have a selective
absorption of polarization components. For instance, this is the reason why the
blue line of the He {\sc i} 10830 \AA\ multiplet shows a non-zero linear
polarization signal in the $B=100$ G right panel of Fig. \ref{fig:fig2}.

\subsection{The weak field regime}

As seen in Fig. \ref{fig:fig2}, for magnetic strengths $B{\lesssim}100$ gauss there is no
significant contribution of the transverse Zeeman effect. In this {\em weak
field} regime the emergent linear polarization is completely dominated by the
atomic polarization of the lower and upper levels of the three line transitions
involved. We emphasize that upper-level polarization leads to a {\em selective emission} of polarization components, while lower-level polarization to a {\em selective absorption} of polarization components
(see Fig. 1).

As pointed out by \cite{trujillo_nature02}, selective absorption is the only
mechanism that can produce linear polarization in the blue component of the He
{\sc i} 10830 \AA\ multiplet. This is because the total angular momentum of its
upper level is $J=0$, which implies that, in the weak field regime under
consideration, $\epsilon_Q=0$. 
The reason why the off-limb panels of Fig. 2
with $B=0$ and $B=100$ gauss show no linear
polarization in that blue line is that our slab with $\Delta{\tau}_{\rm
red}=1$ is optically thin at that blue wavelength, and also because Eqs. (11)
and (12) with $I_0=X_0=0$ imply

\begin{equation}
{\frac{Q}{I}}\,=\,{\frac{\epsilon_Q}{\epsilon_I}},
\end{equation}
which in the weak field regime is zero for the blue component of the helium multiplet (because its upper level, with $J=0$, is intrinsically unpolarizable and $\epsilon_Q=0$).
Interestingly, the same Eqs. (11) and (12) indicate that, if the
boundary intensity $I_0$ were non zero, then there should be a significant $Q/I$
signal. In fact, this is the situation we have in all the right panels of Fig.
\ref{fig:fig2}, which correspond to simulated observations at solar disk center.
 
Obviously, due to symmetry reasons, forward scattering processes in the absence
of a magnetic field produce zero polarization, as seen in the $B=0$ gauss right
panel of Fig. 2. The same applies if there is a microturbulent magnetic field, or if the
magnetic field vector is parallel to the symmetry axis of the radiation field
that illuminates the slab's helium atoms. However, as shown in the $B=100$ gauss
right panel of Fig. 2, forward scattering processes in the presence of an inclined
magnetic field do produce linear polarization in the lines of the He {\sc i} 
10830 \AA\ multiplet. Here, the linear polarization 
is actually {\em created} by the Hanle effect. This is easy to understand by
reasoning within the framework of the oscillator model for the Hanle effect in a
triplet-type transition with $J_l=0$ and $J_u=1$ \citep[e.g.,][]{trujillo01}.
For the more general case of a line transition between two isolated levels
having any possible $J_l$ and $J_u$ values, it is first necessary to recall
\citep[e.g.,][]{landi_landolfi04} that the Hanle effect tends to
destroy the quantum coherences between the magnetic sublevels pertaining to each
$J$-level, {\em without modifying the population imbalances}\footnote{This result is strictly valid for a two-level atomic model 
in the so-called magnetic field reference system, whose z-axis (i.e., the quantization axis for total angular momentum) is aligned
with the magnetic field vector. 
We recall that once a reference system is
chosen, the quantum coherences and the population imbalances can be conveniently
quantified by using the multipole components of the atomic density matrix
corresponding to the $J$-level under consideration, which are commonly denoted
by the symbol ${\rho}^K_Q(J,J^{'})$.}. In the magnetic field reference system, the
quantum coherences (i.e., the ${\rho}^K_Q$ values with $Q{\ne}0$) tend to vanish for
magnetic strengths $B>B_{\rm satur}{\approx}10\,B_H$ (that is, in the saturation
regime of the Hanle effect), so that in practice
in this regime we are only left with the
population imbalances among the magnetic sublevels pertaining to each $J$-level
(i.e., with the ${\rho}^K_Q$ values with $Q=0$). Therefore, in the presence of a
magnetic field inclined with respect to the symmetry axis of the pumping
radiation field, the population imbalances of the upper and/or lower levels can
produce linear polarization parallel or perpendicular to the horizontal
component of the magnetic field vector, simply because of the ensuing selective
emission and selective absorption of polarization components. The amplitude of
the resulting Stokes $Q$ signal can be easily estimated, in the optically thin limit of Eq. (12) and/or in the optically thick limit of Eq. (14), by introducing into such expressions the following approximate formulae (see Trujillo Bueno 2003b):

\begin{equation}
{\frac{\epsilon_Q}{\epsilon_I}}\,{\approx}\,{\frac{3}{2\sqrt{2}}}\,(1-{\mu}_B^2)\,{\cal W}\,\sigma^2_0(J_u)\, ,
\label{eq:QIa}
\end{equation}

\begin{equation}
{\frac{\eta_Q}{\eta_I}}\,{\approx}\,{\frac{3}{2\sqrt{2}}}\,(1-{\mu}_B^2)\,{\cal Z}\,\sigma^2_0(J_l)\, ,
\label{eq:QIb}
\end{equation}
where $\sigma^2_0=\rho^2_0/\rho^0_0$ quantifies the degree of population
imbalance of the $J$-level under consideration, while ${\cal W}$ and ${\cal Z}$
are numerical coefficients that depend on the $J_l$ and $J_u$
values\footnote{Actually, ${\cal W}=w^{(2)}_{J_uJ_l}$ and ${\cal
Z}=w^{(2)}_{J_lJ_u}$, with $w^{(2)}_{JJ^{'}}$ given by Eq. (10.12) of 
\cite{landi_landolfi04} for $K=2$.}
(e.g., ${\cal W}=0$ and ${\cal Z}=1$ for a line transition with $J_l=1$ and
$J_u=0$, ${\cal W}=1$ and ${\cal Z}=0$ for a line transition with $J_l=0$ and
$J_u=1$, and ${\cal W}={\cal Z}=-1/2$ for a line transition with $J_l=J_u=1$).
It is very important to note that in Eqs. (\ref{eq:QIa}) and (\ref{eq:QIb}) 
${\mu}_B={\rm cos}\,{\theta}_B$,
where ${\theta}_B$ is {\em the angle between the magnetic field vector and the line of sight}. 

Consider the case of our slab of chromospheric plasma at a given height above
the visible solar ``surface" and permeated by a horizontal magnetic field (see
Fig. 1).  
A magnetic field of 100 gauss is more than sufficient to seriously reduce the quantum
coherences (as quantified in the magnetic field reference frame), but is still
sufficiently weak so as to be sure that the contribution from the transverse
Zeeman effect is negligible. For the He {\sc i} multiplet this happens for
$10{\lesssim}B{\lesssim}100$ G, approximately.
Under such circumstances the above-mentioned expressions should provide a reasonable approximation for
estimating the emergent $Q/I$ at the line center of a significantly strong
spectral line. In agreement with our detailed numerical calculations such
approximate formulae predict linear polarization for a disk center observation,
be it for a line transition with $J_l=1$ and $J_u=0$, or for one with $J_l=0$
and $J_u=1$, or for one with non-zero $J_l$ and $J_u$ values (except the
particular case of a spectral line with $J_l=J_u=1/2$, because a level with $J=1/2$ cannot be aligned). 

\subsection{The strong field regime}

We now turn our attention to discussing the cases of Fig. \ref{fig:fig2} for which the
transverse Zeeman effect plays a significant role -that is, those with
$B{\gtrsim}500$ gauss. As seen in the corresponding panels, the influence of atomic
level polarization on the emergent linear polarization is significant, mainly
for the red blended component (which results from two transitions whose upper
levels can be polarized). As expected, the stronger the field the smaller the
difference between the solid-line profiles (which show the joint effect of all
physical ingredients) and the dotted profiles (which neglect the influence of
atomic level polarization). The dashed profiles assume that the lower level is
unpolarized, but take into account the selective emission processes that result
from the presence of upper-level atomic polarization. For a slab with
$\Delta{\tau}_{\rm red}=1$ (which implies a smaller optical thickness at the
wavelength of the blue line!) the influence of lower-level polarization is
significant for the $B=500$ gauss disk center case, but insignificant for
sensibly stronger fields. However, upper-level polarization plays an important
role on the linear polarization of the red blended component, 
even for magnetic strengths as large as 1000 gauss, for both the off-limb and
on-disk cases. For example, for the $B=500$ gauss case (which is representative
of the strengths found in emerging flux regions) the accurately computed
emergent Stokes $Q$ profile is very different from that obtained taking into
account only the influence of the transverse Zeeman effect within the framework
of the Paschen-Back effect theory.

\subsection{Observational evidence in emerging flux regions}

Interestingly, the above-mentioned observational signature of the presence of
atomic polarization in the helium levels is clearly seen in many of the Stokes
profiles observed in emerging flux regions, such as those of Fig. 2 of \cite{Lagg04}. 
The solid lines of our 
Fig. \ref{fig:fig3} show a good theoretical fit to such
observations (see Fig. 2 of Lagg et al. 2004), which we have 
obtained by taking into account the influence of atomic level polarization, in
addition to that of the Zeeman effect. The dotted lines of Fig. \ref{fig:fig3} neglect,
however, the influence of atomic level polarization. A comparison of such theoretical Stokes
profiles with those observed by \cite{Lagg04}
indicates the presence of atomic level polarization
in a relatively strong field region (${\sim}$1000 gauss). It also shows that 
neglecting the influence of atomic level polarization
on the emergent Stokes profiles is a suitable approximation for interpreting the
circular polarization, but 
an unsuitable one for modeling the observed linear polarization. We conclude
that the Stokes $Q$ and $U$ profiles observed by \cite{Lagg04} in emerging
flux regions are strongly modified by the presence of atomic level polarization,
even at those atmospheric points of the observed field of view for which field
strengths as large as 1000 gauss are inferred. In any case, it may be
tranquilizing to point out that an inversion of the observed profiles neglecting
atomic level polarization yields a similar magnetic field vector, in spite of
the fact that the corresponding theoretical fit is very poor. A much better
theoretical fit is automatically obtained when the influence of atomic level
polarization is properly taken into account, which is important for the
reliability of the Stokes inversion results (see Asensio Ramos \& Trujillo Bueno 2006, in preparation, for a detailed description of our forward modeling and inversion codes we have applied in this investigation). 

\section{The influence of horizontal illumination}

At the Fourth International Workshop on Solar Polarization (SPW4) that took
place in Boulder (USA) during September 2005, V.
Mart\'\i nez Pillet and collaborators reported on spectropolarimetric
observations of (low-lying) active region filaments 
in the He {\sc i} 10830 \AA\ multiplet, taken with the new version (TIP-2) of
the Tenerife Infrared Polarimeter (Mart\'\i nez Pillet et al. 2006; in
preparation). In particular, in his SPW4 talk V.
Mart\'\i nez Pillet pointed out that the Stokes profiles of the observed active
region filaments had the typical shape of polarization profiles produced by
the Zeeman effect,
without showing any hint at all of the observational
signature of the Hanle effect in forward scattering that \cite{trujillo_nature02} 
had found in solar coronal filaments located at relatively large heights 
above quiet regions of the solar ``surface". In other words, the typical shapes of the linear polarization profiles observed by Mart\'\i nez Pillet et al. in active region filaments were similar to those of the dotted lines of Fig. 3. He also reported that Hanle-effect
signals similar to those found by \cite{trujillo_nature02} were however
present in the quiet regions of the observed field of view, outside the filament plasma. 

Which could be the explanation of this enigmatic observational finding? One possibility is that elastic collisions with neutral hydrogen atoms completely
destroy the polarization of the helium levels, 
but this is very unlikely because the typical densities of solar plasma
structures are too low to affect the (short-lived) upper levels. 
In fact, simple estimates based on Eq. (7.108) of Landi Degl'Innocenti \& Landolfi (2004) suggest that at a height of 2000 km in the FAL-C semi-empirical model of Fontenla et al. (1993) the upper-level rates of depolarizing collisions are about four orders of magnitude smaller than the corresponding Einstein $A_{ul}$-coefficient. Therefore, collisional depolarization seems to be indeed negligible, even if the hydrogen number density of active region filaments were three orders of magnitude larger than those of the FAL-C model at chromospheric heights. Another depolarizing possibility could be the bound-free transitions caused by the UV ionizing radiation coming downwards from the corona, but as is well known most of the ionizations take place from the singlet states of He {\sc i} (instead of from the triplet states of the 10830 \AA\ multiplet) and, in any case, the intensity of the ionizing radiation seems to be too low so as to produce any significant effect (Casini \& Manso Sainz 2006; private communication).  
 
In our opinion, what is happening here is that the radiation field generated by
the active region filament itself 
(which is not an optically-thin structure!) 
makes a negative contribution to the
anisotropy factor, so that the anisotropy of the true radiation field that
illuminates the slab's helium atoms is negligible. In order to investigate this
possibility we have calculated the anisotropy factor at each point inside a
slab of total optical thickness $\tau_{\rm tot}$  in a way similar to that followed by Asensio Ramos, Landi Degl'Innocenti \& Trujillo Bueno (2005), 
but taking into account the center-to-limb variation of the photospheric
radiation field. If the center-to-limb variation is parameterized by the following
functional form \citep{cox00}
\begin{equation}
 \frac{I(\mu)}{I(\mu=1)} = 1-u(1-\mu) - v(1-\mu^2),
\end{equation}
the resulting expression for the anisotropy factor is\footnote{For simplicity, we write down the equation for the zero height case, while our Fig. 4 below is for a slab located at 3 arc seconds above the solar visible ``surface".}.

\begin{equation}
w=\frac{1}{2} \frac{ a+b (I_0/S_0)}{a'+b' (I_0/S_0)},
\end{equation}
where $\tau$ is measured starting at the lower boundary. The quantities $a$,  
$b$, $a'$ and $b'$ are given by:
\begin{eqnarray}
a &=& \exponint{2}{\tau} + \exponint{2}{\tau_\mathrm{tot}-\tau} - 3 \exponint{4}{\tau} - 
3 \exponint{4}{\tau_\mathrm{tot}-\tau} \nonumber \\
b &=& 3 \exponint{4}{\tau} - \exponint{2}{\tau} + \Big[ \exponint{2}{\tau} - \exponint{3}{\tau} -
3 \exponint{4}{\tau} + 3\exponint{5}{\tau} \Big]u + 
\Big[ \exponint{2}{\tau} - 4\exponint{4}{\tau} + 3\exponint{6}{\tau}\Big]v \nonumber \\
a' &=& 2- \exponint{2}{\tau} - \exponint{2}{\tau_\mathrm{tot} - \tau} \nonumber \\
b' &=& \exponint{2}{\tau} - \Big[ \exponint{2}{\tau} - \exponint{3}{\tau} \Big]u - \Big[ \exponint{2}{\tau} 
- \exponint{4}{\tau} \Big]v.
\end{eqnarray}
The quantities $u$ and $v$ are the coefficients of the center-to-limb variation given by \cite{cox00},
while $\exponint{n}{\tau}$ is the exponential integral of order $n$.
It is also interesting to point out that for the case of an isolated slab located at any given height above the solar ``visible" surface the previous expression simplifies considerably, since it is given by

\begin{equation}
w=\frac{1}{2} \frac{ a}{a'} .
\end{equation}

Figure \ref{fig:fig4} shows how the anisotropy factor varies inside a
slab of optical thickness $\tau_{\rm tot}=1$, which is
located at only 3 arc seconds above the visible ``surface" and is illuminated from below by the continuum photospheric radiation.
The results are shown
for increasing values of the ratio $I_0/S_0$ (where $I_0=I(\mu=1)$ is the
intensity of the vertical ray coming from the underlying photosphere, while
$S_0=S_I$ is the slab's source function).
The dotted horizontal line indicates the value of the anisotropy factor, $w$, that we would
have if the contribution of the radiation field generated by the slab itself
were neglected. The remaining curves show the anisotropy factor for a range of
$I_0/S_0$ values. As we can see, there is a range of $I_0/S_0$ values around
$I_0/S_0=1$ (i.e., the expected values when the atomic excitation is dominated by radiative transitions)
for which $w{\approx}0$ at many points inside the slab. In our
opinion, these are precisely the 
physical conditions inside the active region filaments observed by V. Mart\'\i nez Pillet et al., which are 
elongated plasma structures located at relatively low heights above the visible solar ``surface".

The illumination conditions in the emerging flux regions observed by \cite{Lagg04} 
are clearly different. Here the illumination is due to  the typical
radiation field of a stratified atmosphere, whose anisotropy is dominated by the positive contribution caused by the
limb darkening of the outgoing radiation (e.g., Trujillo Bueno 2001). On the contrary,
active region filaments are elongated plasma structures located at relatively
low heights above the solar visible ``surface". Such plasma structures ``levitating" in
the solar atmosphere have a significant optical thickness, which implies that we
have to take into account the negative contribution to $w$ caused by the
(mainly) horizontal radiation field generated by the filament itself.  
If this reduction of the radiation field's anisotropy is not taken into account when doing inversions of the Stokes $Q$ and $U$ profiles observed in such type of optically-thick structures (that is, when only the positive anisotropy of the photospheric radiation field is considered), the inversion algorithm may artificially select magnetic field vectors with inclinations around the Van Vleck angle ($\theta_B=54.74^{\circ}$), because it is for this particular inclination that the contribution of atomic level polarization is minimized. Fortunately, the observed Stokes $V$ profile is often available, but it is very unlikely that the observed linear polarization is automatically reproduced after choosing a magnetic strength that fits the observed circular polarization assuming that the magnetic field inclination is close to that of the Van Vleck angle. Obviously, when a spectropolarimetric observation of the He {\sc i} 10830 \AA\ multiplet does not show any hint at all of the presence of atomic level polarization,
the best one can do for inferring the magnetic field vector is to apply an inversion code which takes only into account the Zeeman effect, but with the positions and strengths of the $\pi$ and $\sigma$ components calculated within the framework of the Paschen-Back effect theory.

\section{Conclusions}

Probably, one of our most interesting conclusions is that the modeling of the emergent Stokes $Q$ and $U$ profiles of the He {\sc i} 10830 \AA\ multiplet should not be done by neglecting the possible presence of atomic level polarization. Actually, the degree of anisotropy of the solar continuum radiation at 10830 \AA\ is sufficiently important so as to produce a sizable amount of atomic alignment in the lower and upper levels of the He {\sc i} 10830 \AA\ multiplet, even at relatively low atmospheric heights such as 1000 Km.

For weak magnetic fields (e.g., $B{\lesssim}100$ G) the emergent linear polarization is fully dominated by the {\em selective emission} and {\em selective absorption} of polarization components that result from this atomic level polarization. For stronger magnetic fields, the contribution of the transverse Zeeman effect cannot be neglected. However, the emergent linear polarization may still show an important 
contribution caused by the presence of atomic level polarization, even for magnetic strengths as large as 1000 G (see Fig. 2).

In emerging magnetic flux regions the anisotropic illumination of the helium atoms is expected to be more or less similar to that corresponding to a stratified stellar atmosphere, with the (mainly vertical) outgoing radiation contributing with positive values to the anisotropy factor, $w$, and with the (mainly horizontal) incoming radiation contributing with negative values. As a result, $w{\approx}0.097$ at a height of 3 arc seconds in the quiet solar atmosphere and a significant amount of lower-level and upper-level polarization is induced by the ensuing radiative transitions. Interestingly, the observational signature of this atomic level polarization is clearly seen in many of the Stokes $Q$ and $U$ profiles observed by Lagg et al. (2004), even at points of the observed field of view for which magnetic strengths as large as 1000 G are inferred.

The illumination conditions are however different in optically thick plasma structures embedded in the solar atmosphere, such as those encountered in (low lying) active regions filaments. Here, in addition to the positive contribution to $w$ due to the continuum radiation field coming from the underlying solar photosphere, we have to take into account the negative contribution caused by the radiation field generated by the optically-thick plasma structure itself.  As a result, the anisotropy factor of the true radiation field that illuminates the helium atoms inside the filament plasma may be very different, as illustrated in Fig. 4 for the case of a constant-property slab of total optical thickness unity. Interestingly, for $I_0/S_0{\sim 1}$ we find that $w{\approx}0$ at many points inside the slab (with $I_0$ the intensity of the vertical ray of the photospheric radiation and $S_0$ the slab's source function). Under such circumstances the ensuing atomic level polarization turns out to be negligible and the emergent Stokes $Q$ and $U$ profiles are dominated by the transverse Zeeman effect. In our opinion, this is the main reason that explains the absence of any observational signature of the presence of atomic level polarization in the linear polarization profiles observed by Mart\'\i nez Pillet and collaborators in active region filaments.

Finally, it is interesting to point out that for the case of isolated plasma structures located at any given height above the solar visible ``surface" the anisotropy factor may reach a significant value (see the thin solid-line curve of Fig. 4, which corresponds to $I_0/S_0{=}0$). This suggests that a variety of plasma structures, like loops embedded in the $10^6$ K solar corona, should produce measurable linear polarization even at spectral line wavelengths for which the underlying solar disk is seen completely dark (e.g., at the EUV and X-ray spectral regions). This last point emphasizes further the relevance of the scientific case for spectropolarimetry from space, so that space agencies worldwide should indeed try to open soon this new diagnostic window on the Universe.\footnote{See Trujillo Bueno et al. (2005b) for additional arguments on the scientific case for spectropolarimetry from space.}

\acknowledgments 
We would like to thank Roberto Casini (HAO), Rafael Manso Sainz (IAC) and Egidio Landi Degl'Innocenti (Universit\`a di Firenze) for stimulating scientific discussions and for their careful reading of this paper. The content of Section 4 was motivated by the interesting observational results on active region filaments presented by V. Mart\'\i nez Pillet and collaborators during the Fourth International Workshop on Solar Polarization, which took place in Boulder (USA) during September 2005. This paper has been completed during 
the Summer of 2006 while JTB was holding the {\em Gauss-Professur} granted by the {\em Akademie der Wissenschaften zu G\"ottingen}. He  wishes to thank Franz Kneer and the rest of the colleagues of the {\em Institut f\"ur Astrophysik} for their hospitality and interest on this type of investigations on the physics of scattering polarization.
This research has been partially funded by the Spanish Ministerio de Educaci\'on y Ciencia through project AYA2004-05792.


\begin{figure}
\plottwo{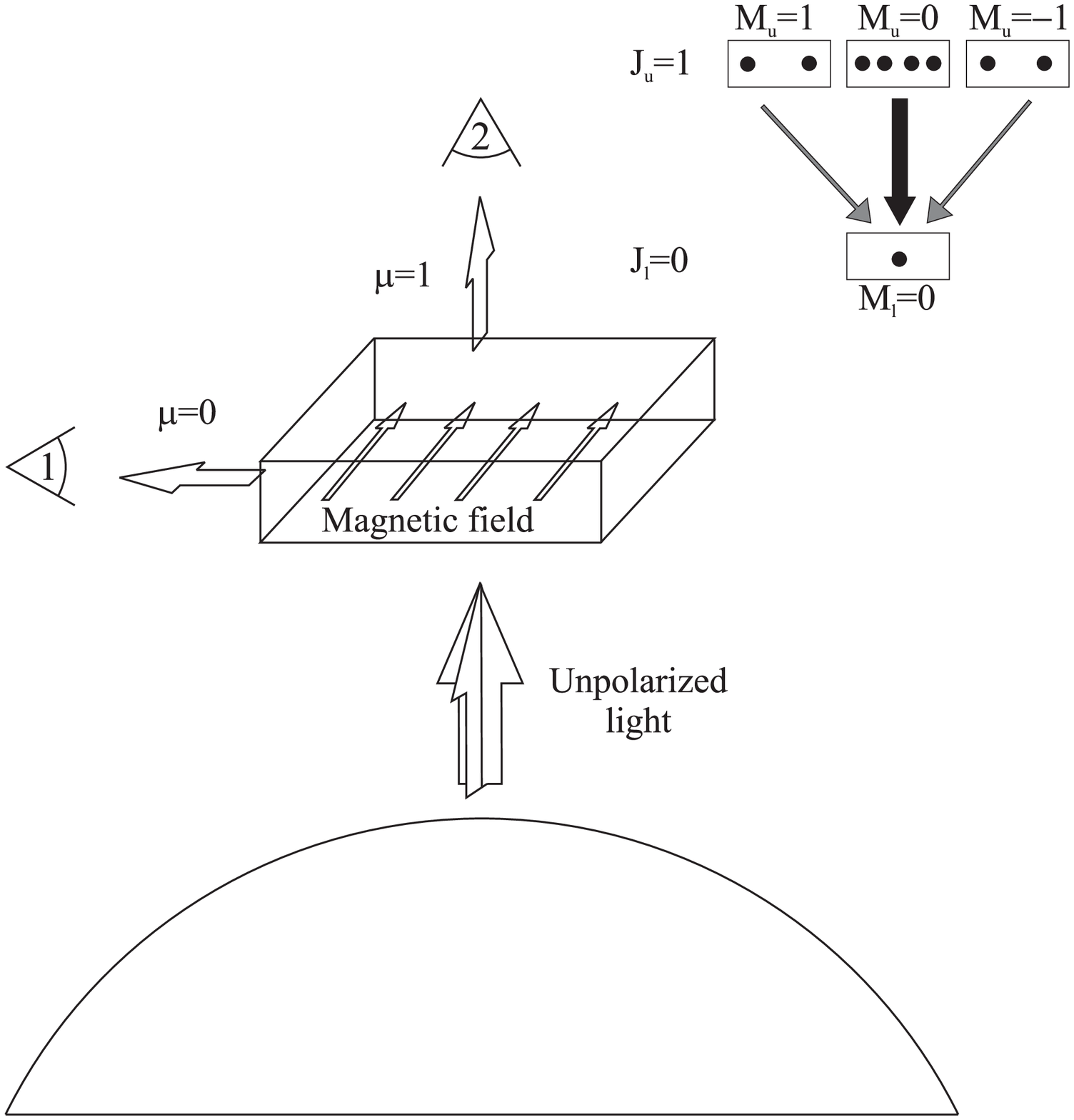}{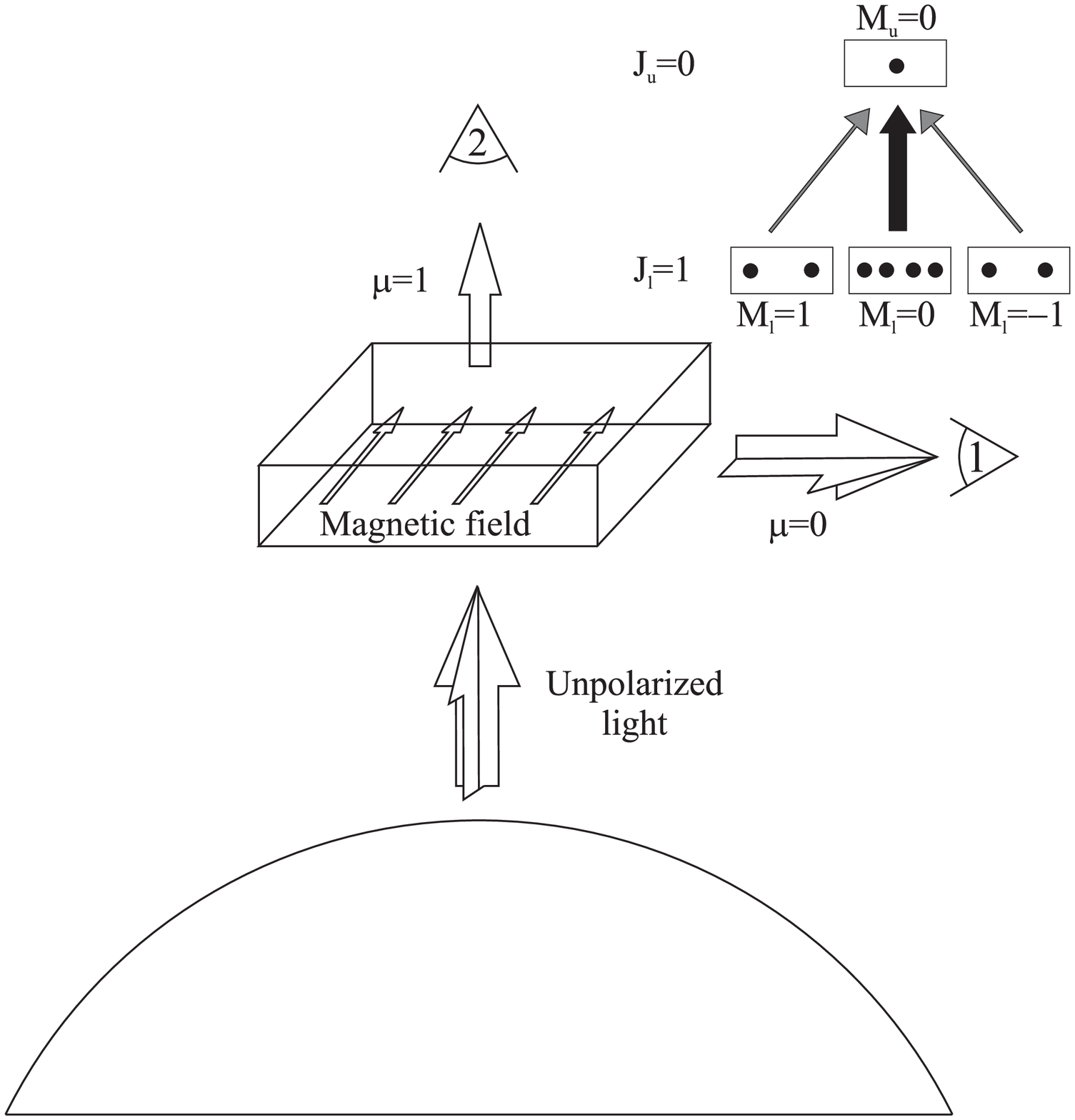}
\caption{Illustration of the emergent polarization that results
from $90^\circ$ and from forward scattering events in the presence of a magnetic field parallel to the solar surface, which in this figure is assumed to be weak but still
sufficiently strong so as to destroy any quantum coherences in the magnetic field reference frame (i.e., $10{\lesssim}B{\lesssim}100$ G).
The left panel refers to a resonance line with $J_l=0$ and $J_u=1$, where we have assumed that the population imbalances are established by the resonance line radiation itself. The right panel refers to the ``blue" line of the He {\sc i} 10830 \AA\ multiplet, which has $J_l=1$ and $J_u=0$, and for which the lower-level polarization is influenced by repopulation pumping (that is, by the spontaneous transitions from the polarized upper levels, with $J=2$ and $J=1$, of the He {\sc i} 10830 \AA\ multiplet; see Trujillo Bueno et al. 2002).
Therefore, in the left panel the observed
linear polarization is caused by {\em selective emission},
while in the right panel the only mechanism that can produce
linear polarization is {\em selective absorption}. For this
reason the observer at position ``1'' in the r.h.s. panel
sees that the light scattered at $90^\circ$ by the  
weakly-magnetized and optically-thin plasma is virtually unpolarized.}
\label{fig:fig1}
\end{figure}

\begin{figure}
\begin{center}
\includegraphics[height=160mm]{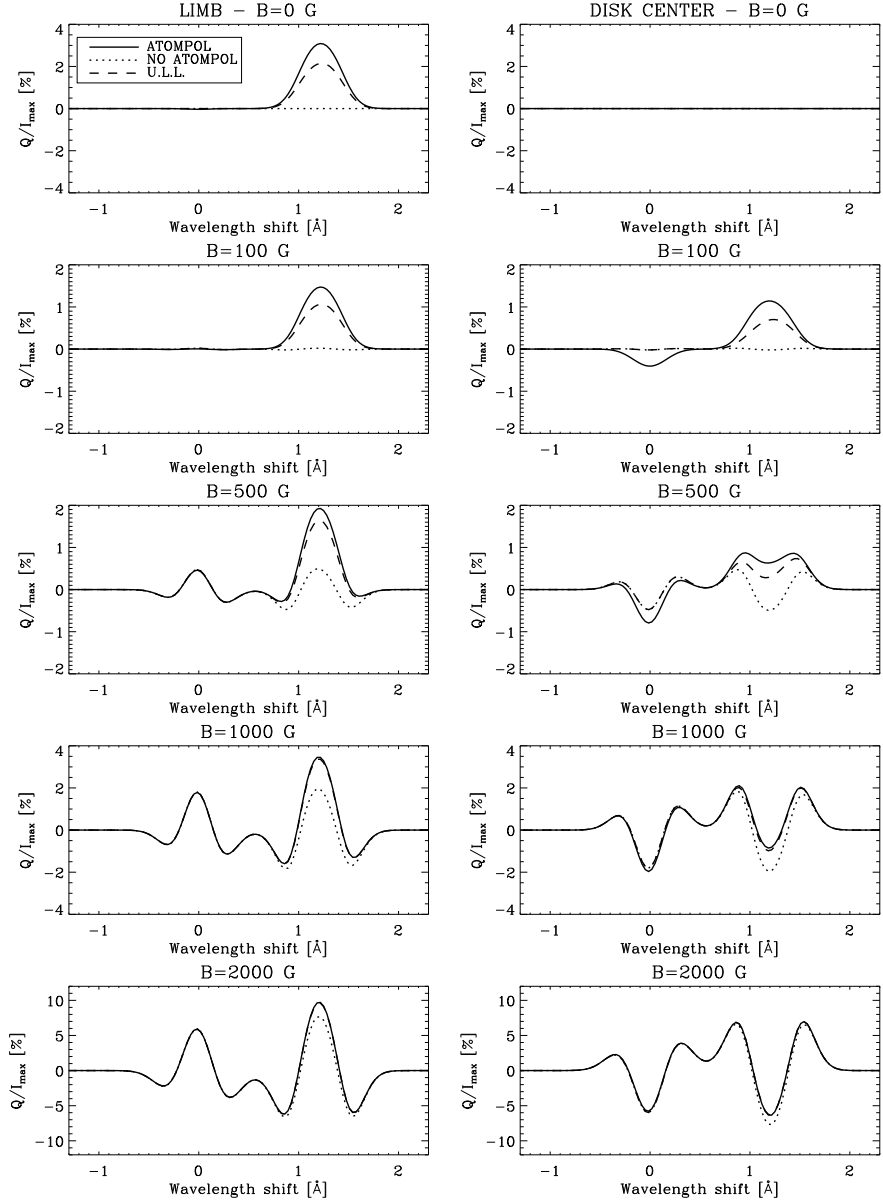}
\end{center}
\caption{The emergent Stokes $Q$ profiles of the 
He {\sc i} 10830 \AA\ multiplet calculated for the two line of sights illustrated in Fig. 1: 
$90^{\circ}$ scattering (left panels, where the Stokes $Q$ profile is normalized to the maximum line-core intensity of the Stokes $I$ emission profile of the `red' line) and forward scattering (right panels, where the Stokes $Q$ profile is normalized to the maximum line-core depression of the Stokes $I$ absorption profile of the `red' line)). Each panel shows the results of three possible calculations for a given strength of the assumed horizontal magnetic field, whose orientation is as shown in Fig. 1 --that is, such that the magnetic field vector is always perpendicular to the line of sight. While the dotted lines neglect the influence of atomic level polarization, the solid lines take it fully into account. The dashed lines show what happens when only the lower level of the multiplet is assumed to be completely unpolarized. The calculations have been carried out for a thermal velocity $v_{\rm T}=6.5\,{\rm km s^{-1}}$.
The positive reference direction for Stokes $Q$ is along the direction of the horizontal magnetic field.}
\label{fig:fig2}
\end{figure}

\begin{figure}
\plotone{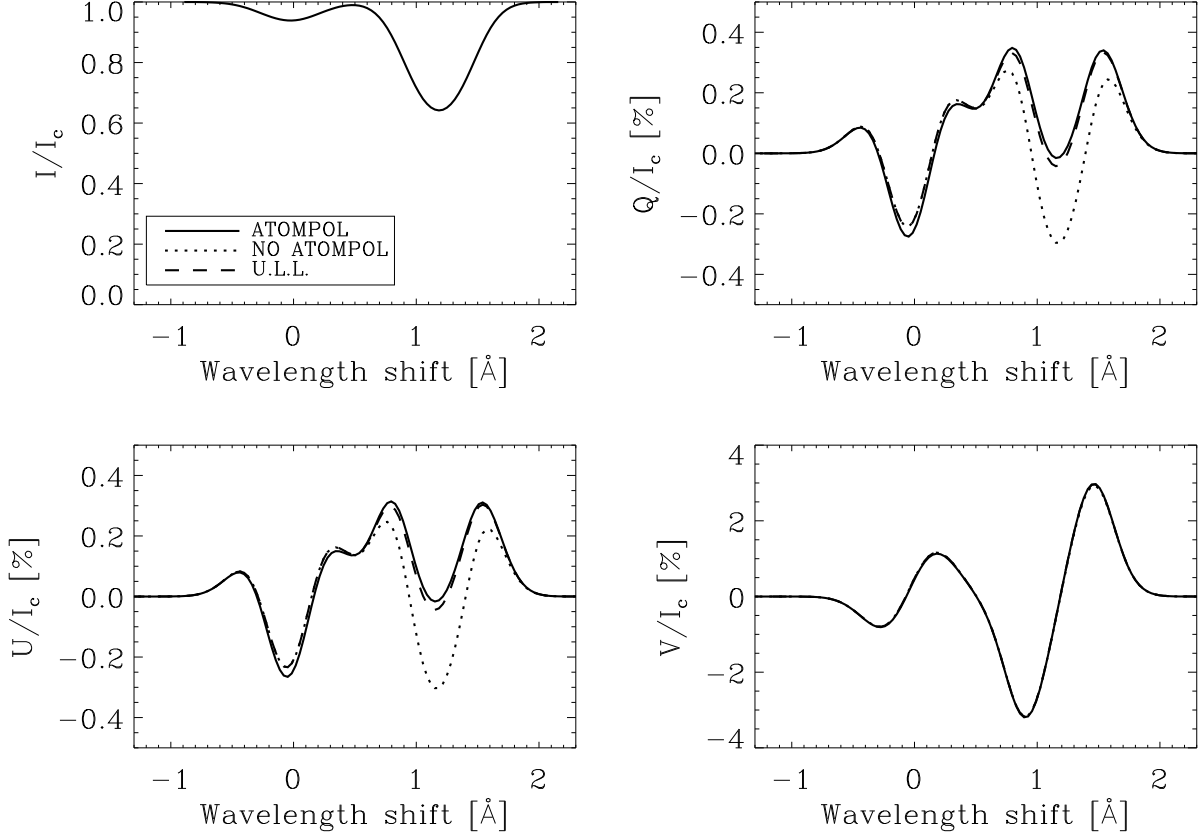}
\caption{The emergent Stokes profiles in the He {\sc i} 10830 \AA\ 
multiplet calculated for the forward scattering case using the following parameters that resulted from the application of our Stokes inversion code: $\Delta{\tau}_{\rm red}=0.87$, thermal velocity ${v}_{\rm T}=8.36\, {\rm km s^{-1}}$, atmospheric height $h=3^{"}$, magnetic strength $B=1070$ G, inclination $\theta_B=86^{\circ}$ and azimuth $\chi_B=-160^{\circ}$. The positive reference direction for Stokes $Q$ is along the x-axis with respect to which the magnetic field azimuth, $\chi_B$, is measured. The Stokes profiles are normalized to the local continuum intensity.
While the dotted lines neglect the influence of atomic level polarization, the solid lines take it fully into account. The dashed lines show what happens when only the lower level of the multiplet is assumed to be completely unpolarized. We point out that, within the framework of our modeling approach, the best fit to the observations shown by Lagg et al. (2004) in their Fig. 2 is provided by the solid lines.}
\label{fig:fig3}
\end{figure}

\begin{figure}
\plotone{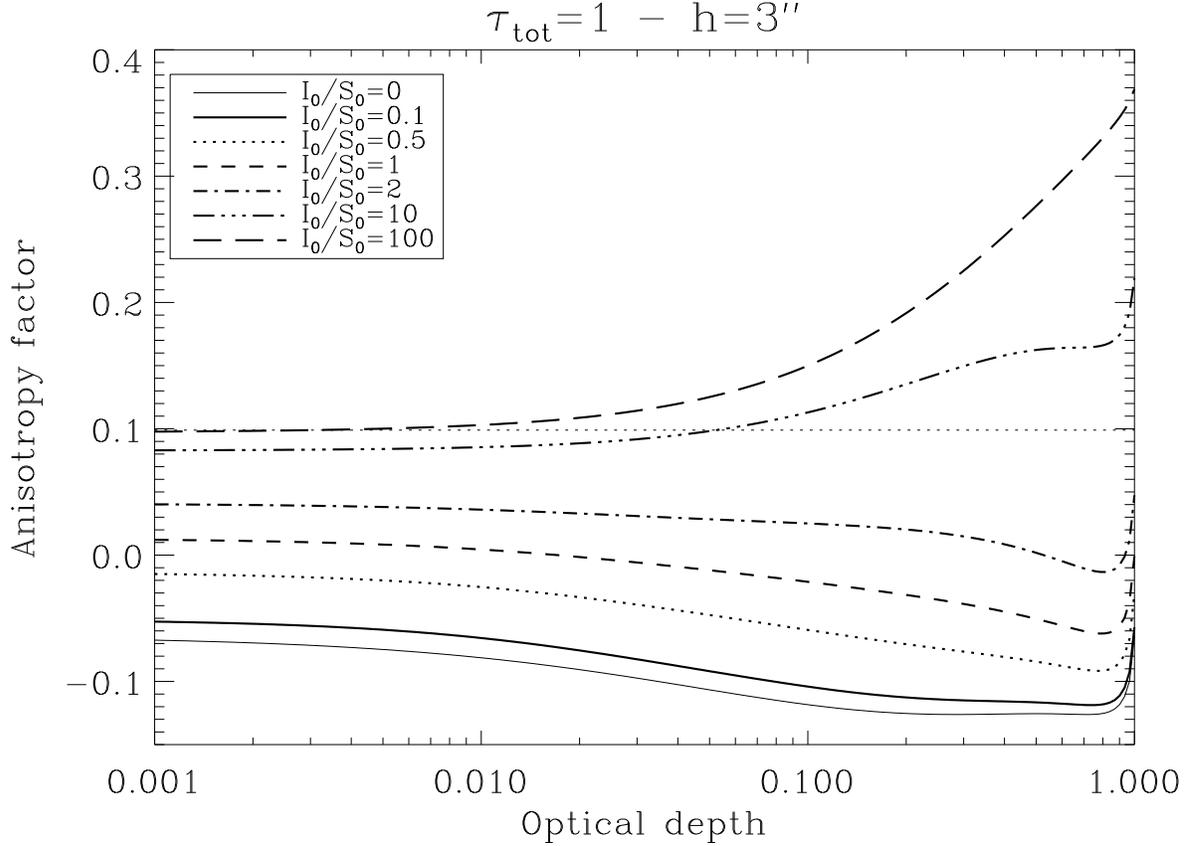}
\caption{Variation of the anisotropy factor (see Eq. 2) 
inside a slab of total optical depth unity and source function $S_0$ located at a height of 3 arc seconds above the solar visible ``surface". The lower boundary of the slab, from where the optical depth along the vertical direction is measured, is illuminated from below by the solar continuum radiation at 10830 \AA, whose intensity at $\mu=1$ is $I_0$. Each curve corresponds to the $I_0/S_0$ value indicated in the inset. Note that for $I_0/S_0{\approx}1$ (i.e., the expected value when the atomic excitation is dominated by radiative transitions)
the anisotropy factor takes very small values at many points inside the slab. The dotted line indicates the value of the anisotropy factor corresponding to the case in which the contribution of the radiation field generated by the slab itself is neglected. Note that for the case of an isolated slab (see the thin solid-line curve corresponding to $I_0/S_0{=}0$) the anisotropy factor is negative and significant (actually, it is even more significant for slabs of smaller optical thickness).}
\label{fig:fig4}
\end{figure}

\end{document}